\documentclass[prc,showpacs,nobibnotes]{revtex4}
\begin{document}

\title{Effect of nucleon exchange on projectile multifragmentation 
in the reactions of $^{28}$Si + $^{112}$Sn and $^{124}$Sn
at 30 and 50 MeV/nucleon.}

\author{M. Veselsky}
\email{veselsky@comp.tamu.edu}
\thanks{On leave of absence from Institute of Physics of SASc, Bratislava, 
Slovakia}
\author{R. W. Ibbotson}
\thanks{Present address: Brookhaven National Laboratory, Brookhaven, NY 11973}
\author{R. Laforest}
\thanks{Present address: Mallinckrodt Institute of Radiology, 
St. Louis, MO 63110}
\author{E. Ramakrishnan}
\thanks{Present address: Microcal Software Inc, One Roundhouse Plaza, 
Northampton, MA 01060}
\author{D. J. Rowland}
\author{A. Ruangma}
\author{E. M. Winchester}
\author{E. Martin}
\author{S. J. Yennello}
\affiliation{Cyclotron Institute, Texas A\&M University, 
College Station, TX 77843.}

\begin{abstract}
{ The multifragmentation of quasiprojectiles was studied in the reactions 
of a $^{28}$Si beam 
with $^{112}$Sn and $^{124}$Sn targets at projectile energies of 30 
and 50 MeV/nucleon. The quasiprojectile observables were 
reconstructed using isotopically identified charged particles 
with $Z_{f} \le $ 5 detected at forward 
angles. The nucleon exchange between projectile and target was investigated 
using the isospin and the excitation energy 
of the reconstructed quasiprojectile. 
For events with total reconstructed charge equal to the charge 
of the beam ( $Z_{tot} =$ 14 ), the influence of the beam energy 
and target isospin on the neutron transfer was studied in detail. 
Simulations were carried out employing a model of deep inelastic transfer, 
a statistical model of multifragmentation and a software replica of the FAUST 
detector array. The concept of deep inelastic transfer 
provides a good description of the production of highly excited 
quasiprojectiles. 
The isospin and excitation energy of the quasiprojectile were described 
with good overall agreement. The fragment multiplicity, charge and 
isospin were reproduced satisfactorily. 
The range of contributing impact parameters was determined using 
a backtracing procedure.}
\end{abstract}


\maketitle

\section*{Introduction~}

Projectile fragmentation has traditionally been thought of as a two-step 
reaction with excitation via a peripheral collision with the target 
followed by fragmentation of the projectile. In this framework, 
the influence of the mass 
and charge of the target nucleus on projectile fragmentation is a 
question of interest both with regard to the formation of the excited 
quasiprojectile and its subsequent fragmentation. The target nucleus 
may affect fragmentation of the projectile in various 
ways. Charity et al. report the influence of the repulsive Coulomb field 
of the target on the motion of the emitted charged particles
\cite{charity}. Additionally, the influence of isospin equilibration on 
reaction dynamics has been studied 
at lower energies \cite{desouza,desouza2,planeta}.
De Souza et al. \cite{desouza} showed that
the nucleon exchange is regulated by the potential energy surface 
if isospin equilibration is allowed to occur. 
At low energies below 10 MeV/nucleon, the studies of nucleon transfer 
\cite{desouza2,planeta} showed deviations from the predictions of the 
commonly used model of nucleon exchange \cite{randrup} in the description 
of a proton and neutron drift. 
At intermediate energies up to 50 MeV/nucleon, the model of nucleon 
exchange successfully describes the production 
of the projectile-like nuclei at forward angles \cite{tassangot}. 
The influence of isospin on the cooling of 
the interacting system by emission of fast nucleons was observed 
in the study of multifragmentation 
of the systems $^{112}$Sn + $^{112}$Sn and $^{124}$Sn~ + $^{124}$Sn 
at a broad range of impact parameters \cite{kunde}. 

In the current study we present a continuation of our previous work 
on projectile multifragmentation of a $^{28}$Si beam in the reaction 
with $^{112}$Sn and $^{124}$Sn targets at 30 and 50 MeV/nucleon 
\cite{laforest}.  
We select the events consisting of isotopically identified
fragments in order to reconstruct the mass and charge of the 
fragmenting projectile-like nucleus. The $N/Z$ difference of the Sn isotopes 
used as targets is significant enough to permit the study of the influence 
of neutron excess on production and deexcitation of the projectile-like 
nucleus. The study is divided into several sections. 
We present a short description of the experimental setup, a discussion of the 
nucleon exchange ( dissipation ) mechanism, divided into an analysis 
of the experimental observables of the reconstructed quasiprojectile and 
a comparison to the results of simulations, 
and a discussion of multifragmentation of excited quasiprojectiles.
Finally a short summary will be presented.

\section*{Experiment}

The experiment was done with a beam of $^{28}$Si impinging on $\sim  $1
mg/cm$^{2}$ self supporting $^{112,124}$Sn targets. The beam was
delivered at 30 and 50 MeV/nucleon by the K500 superconducting cyclotron at
the Cyclotron Institute of Texas A\&M University. The detector array 
FAUST \cite{FAUST} consisted of 68 silicon - CsI(Tl) telescopes covering 
polar angles from 2.3$^{\circ}$ to 33.6$^{\circ}$ in the laboratory system. 
Each element is composed of a 300$ \mu  $m surface barrier silicon detector 
followed by a 3cm CsI(Tl) crystal. The detectors are arranged in five 
concentric rings. The geometrical efficiency is approximately 90$ \%$ 
for the angle range covered.  These detectors allow isotopic identification 
of light charged particles and intermediate-mass fragments up to a charge 
of $ Z_{f}=5 $. The energy thresholds are determined by the energy needed 
to punch through the 300$ \mu$m silicon detector. These energy thresholds 
have little effect on the acceptance of particles from the fragmenting 
projectile due to the boost from the beam energy. Details of the experimental 
procedure and detector calibration can be found in ref. \cite{laforest}. 
Additional silicon telescopes complemented the forward array in the setup. A
telescope consisting of a 53$\mu $m silicon detector, 
147$\mu $m silicon strip detector (16 strips) and 
a 994$\mu $m silicon detector was placed at 40$^{\circ }$ in the
laboratory. The 53$\mu $m and 994$\mu $m silicons had an active area of 5cm $
\times $ 5cm and were divided in four quadrants. This telescope covered the
polar angle from 42.5$^{\circ }$ to 82.2$^{\circ }$ degrees. Another silicon
telescope was placed at 135$^{\circ }$ in the lab, covering polar angles from
123$^{\circ }$ to 147$^{\circ }$. It was composed of two 5cm$\times $5cm
active area silicon detectors of thickness 135$\mu $m and 993$\mu $m
respectively. A 2cm thick CsI(Tl) detector read out via a photo-diode 
was placed behind both silicon pairs. 

In the present study we restrict ourselves to the events where 
all emitted fragments are isotopically identified ( $ Z_{f}<5 $ ). We assume 
that such events detected in the FAUST detector array originate predominantly 
from the deexcitation of the quasiprojectile ( or projectile-like 
source ). The total charge of the reconstructed quasiprojectile ( QP ) 
is restricted to the values near the projectile charge ( $ Z_{tot}=12-15 $ ). 
This very selective data contains information on fragmentation 
of highly excited projectile-like prefragments, and thus can be used for 
the study of the mechanism of dissipation of the kinetic energy of relative 
motion into thermal degrees of freedom.  
The high granularity of FAUST, the moderate beam current, 
and the high selectivity of the events allowed us to minimize 
the number of pile-up signals.

\section*{Nucleon exchange}

Nucleon exchange is supposed to be a highly effective mechanism 
of dissipation of the kinetic energy of relative motion of the projectile 
and target into their internal degrees of freedom. In this section 
we present an overview of experimental observables of the reconstructed 
quasiprojectile and a comparison to the results of simulations.

\subsection*{Experimental observables}

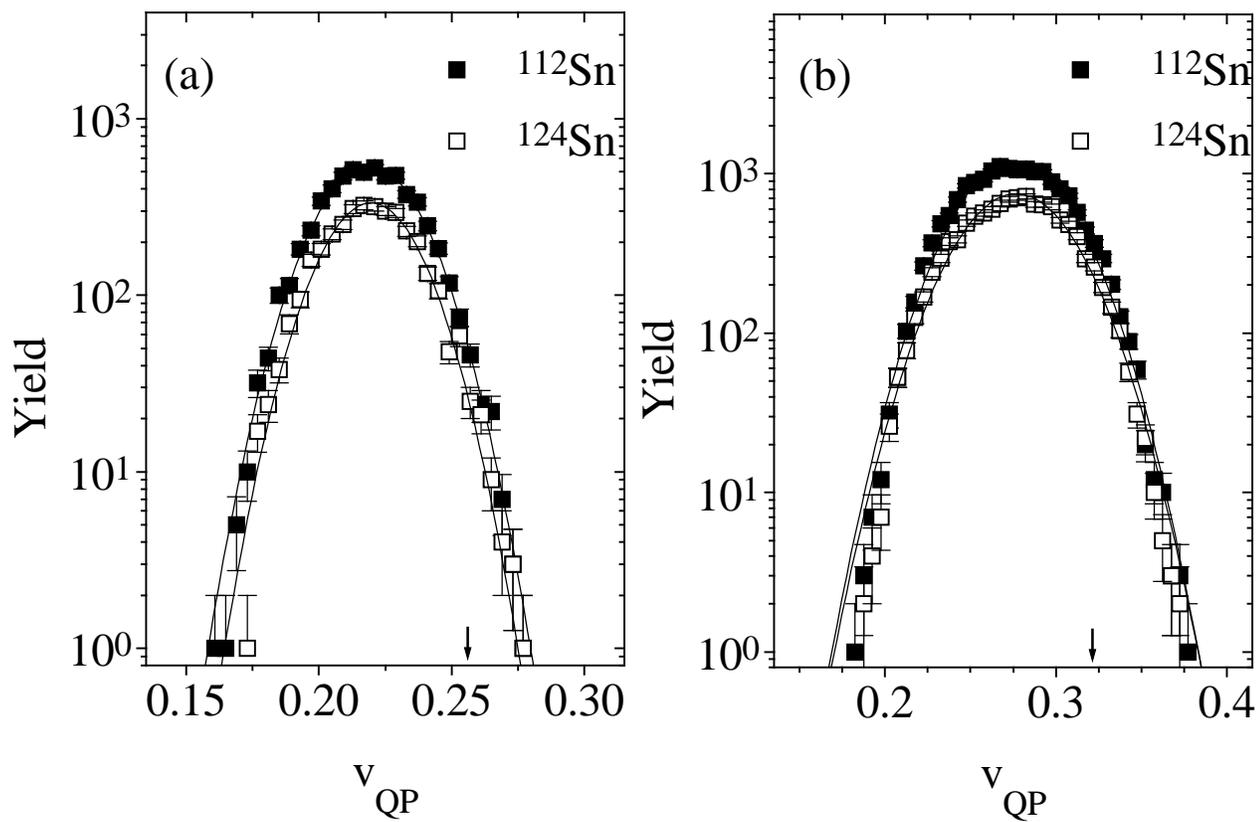
\begin{figure}[!]
\caption{ Experimental velocity distributions of the fully isotopically
resolved quasiprojectiles with $ Z_{tot}=12-15 $ ( 
solid and open squares mean  
$^{112}$Sn and $^{124}$Sn target, 
(a) and (b) the projectile energy 
30 and 50 MeV/nucleon, respectively ).  Solid lines show Gaussian fits.  }
\label{vqpex}
\end{figure}

In order to identify an emitting source 
from which the detected fragments originate, we reconstructed the 
velocity distributions of the quasiprojectiles with total charge 
$ Z_{tot}=12-15 $ for the set of events  where 
all emitted fragments are isotopically identified. 
Resulting velocity distributions for projectile energies 
30 and 50 MeV/nucleon are given on Fig. \ref{vqpex}a,b. 
Solid squares represent the reaction with $^{112}$Sn target
and open squares reaction with $^{124}$Sn. For a given 
projectile energy, the mean velocities and widths of distributions 
are practically identical for both targets. The velocity 
distributions are close to Gaussians over two orders of magnitude 
( Gaussian fits are given as solid lines ). 
The observed velocity distributions are symmetric and have 
no significant low or high energy tails.  
Thus, the reconstructed quasiprojectiles may indeed be identified 
with the projectile-like fragment source. 
The admixture of particles from the midvelocity sources such as 
preequilibrium or neck emission, if any, does not distort the Gaussian 
shape of the quasiprojectile velocity distributions. 
The mean velocities of the sources are somewhat 
lower than the velocity of the beam ( indicated by arrows ), which indicates 
the damping of the kinetic energy into internal degrees of freedom.  

Useful experimental information about the nucleon exchange rate can be found
in the events where the charge of the reconstructed quasiprojectile is equal 
to the charge of incident beam ( $ Z_{tot}=14 $ ). In this case, isospin 
equilibration may only occur by the transfer of neutrons, as the number 
of transferred neutrons is the only available isospin degree of freedom 
of the system. Since the neutron number of the reconstructed 
quasiprojectile is just the sum of neutrons bound in the fragments 
with non-zero charge, we define the principal 
neutron exchange observable as the mass change. 
Subtracting the sum of the neutrons bound in 
detected fragments from the neutron number of the beam gives 

\begin{equation}
 \Delta A = N_{proj}-\sum_{f} N_{f} 
\label{dela}
\end{equation}

where $N_{proj} =$ 14 for $ ^{28} $Si beam. 
A positive value of this observable means that the neutron number of 
the reconstructed quasiprojectile is lower than the neutron number of the 
projectile and one or more projectile neutrons have been lost by transfer to 
the target nucleus and/or by emission in the fragmentation stage. 
A positive value of $ \Delta A$ 
may also be obtained in a collision where the transfer 
of one or more neutrons from the target to the projectile occurs but 
a larger number of neutrons is emitted later.   
Finally, a negative value of $ \Delta A$ means that the neutron 
flow from target to projectile is stronger than emission.   

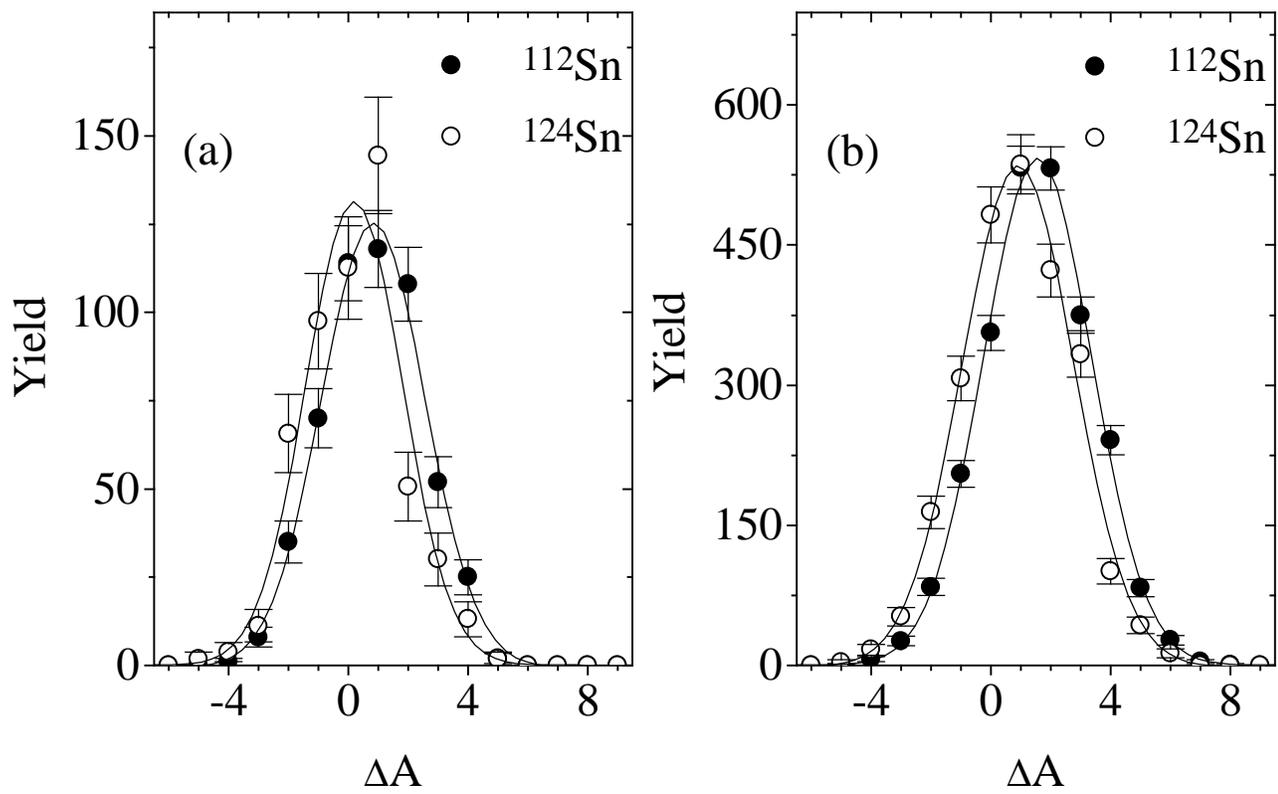
\begin{figure}[!]
\caption{ Experimental distributions of the mass change 
for fully isotopically
resolved quasiprojectiles with $ Z_{tot}=14 $ ( solid and open circles mean  
$^{112}$Sn and $^{124}$Sn target, 
(a) and (b) the projectile energy 
30 and 50 MeV/nucleon, respectively ). 
Solid lines show Gaussian fits.  }
\label{munpex}
\end{figure}

The resulting mass change distributions for both 
projectile energies and target isotopes are shown in Fig. \ref{munpex} 
( $^{112}$Sn - solid circles, $^{124}$Sn - open circles ). 
The mass change depends on the target 
nucleus and beam energy. For both projectile energies the mean value of the 
mass change is larger for the reaction with the $^{112}$Sn 
target by a little more than half a unit  
( 0.60 for 30 MeV/nucleon and 0.65 for 50 MeV/nucleon, see Table 
\ref{table1} ). Therefore, there could be more neutrons transferred 
from the target to the projectile during the interaction with $^{124}$Sn 
target, or there could be more neutrons emitted from the 
quasiprojectile that interacted with the $^{112}$Sn target, 
or more neutrons could be transferred from the projectile to 
the $^{112}$Sn target. The relative importance of different 
processes may be deduced from the sign of the mean values of the $\Delta A$ 
distributions. As one can see in Table \ref{table1}, they are positive in 
all cases. 
For the events with $ Z_{tot}=14 $, where the proton degree of freedom 
is fixed, the only possible way to achieve isospin equilibrium 
of the interacting dinuclear system is with neutron flow from the target 
to the projectile. Thus, especially in the case 
of neutron rich $^{124}$Sn target, the positive values of $<\Delta A>$ 
provide evidence for the influence of the neutron emission 
on the final neutron content of the quasiprojectile. 
Indeed, when comparing the mean values of the mass change 
for the reactions with the same target nucleus at different projectile 
energies, the mean mass change significantly increases
with an increase of beam energy from 30 to 50 MeV/nucleon 
\hbox{( 0.73} for $^{112}$Sn and 0.68 for $^{124}$Sn ). 

\begin{table}[tbp]
\caption{ Mean values and widths of experimental distributions  
of the mass change \hbox{( $\Delta A$ )}  
and of the apparent quasiprojectile excitation energy  
( $E_{app}^{*}$ ) for the fully isotopically resolved quasiprojectiles. }
\label{table1}

{\centering \begin{tabular}{cccccccc}
\hline 
\hline 
&
&
\multicolumn{4}{c}{$ Z_{tot}= $ 14 }&
\multicolumn{2}{c}{ $ Z_{tot}= $ 12 - 15}\\
$ E_{proj} $&
Target&
$ <\Delta A> $ &
$ \sigma _{<\Delta A>} $ &
$ <E^{*}_{app}> $&
$ \sigma _{<E^{*}_{app}>} $&
$ <E^{*}_{app}> $&
$ \sigma _{<E^{*}_{app}>} $\\
(MeV/nucleon)&
&
&
&
(MeV)&
(MeV)&
(MeV)&
(MeV)\\
\hline 
30&
$ ^{112} $Sn&
0.81&
1.63&
101.2&
25.8&
114.9&
25.0\\
&
$ ^{124} $Sn&
0.21&
1.63&
102.3&
27.0&
120.2&
28.3\\
50&
$ ^{112} $Sn&
1.54&
1.82&
142.8&
37.9&
160.8&
36.7\\
&
$ ^{124} $Sn&
0.89&
1.82&
142.8&
38.1&
161.2&
38.1\\
\hline 
\hline 
\end{tabular}\par}

\end{table}

When looking at the shapes of the observed $ \Delta A$ distributions, 
it is apparent that they are almost identical for different targets 
at the same projectile energy and are close to ideal Gaussians. 
The relative shift between systems is given by the mean values discussed 
above. The width of the $ \Delta A$ distributions increases slightly 
( see Table \ref{table1} ) with increasing projectile energy. 
The Gaussian shapes of the $ \Delta A$  distributions may 
be understood as an additional argument for the presence  
of the nucleon exchange in the early stage of the reaction because  
they resemble the predictions of the theory 
of deep inelastic transfer ( DIT ) \cite{randrup,feldmeier}. 
Within this theoretical concept, Gaussian shapes of the mass 
and charge distributions are obtained as solutions 
of transport equations ( e.g. Fokker-Planck ) for 
the mass and charge degrees of freedom.  
Thus, the experimental mean values and shapes of the mass change 
distributions suggest the general 
picture where the mass change is a combination of the number 
of neutrons transferred between projectile and target 
during the interaction phase of the reaction 
and of the number of neutrons emitted from the excited quasiprojectile.

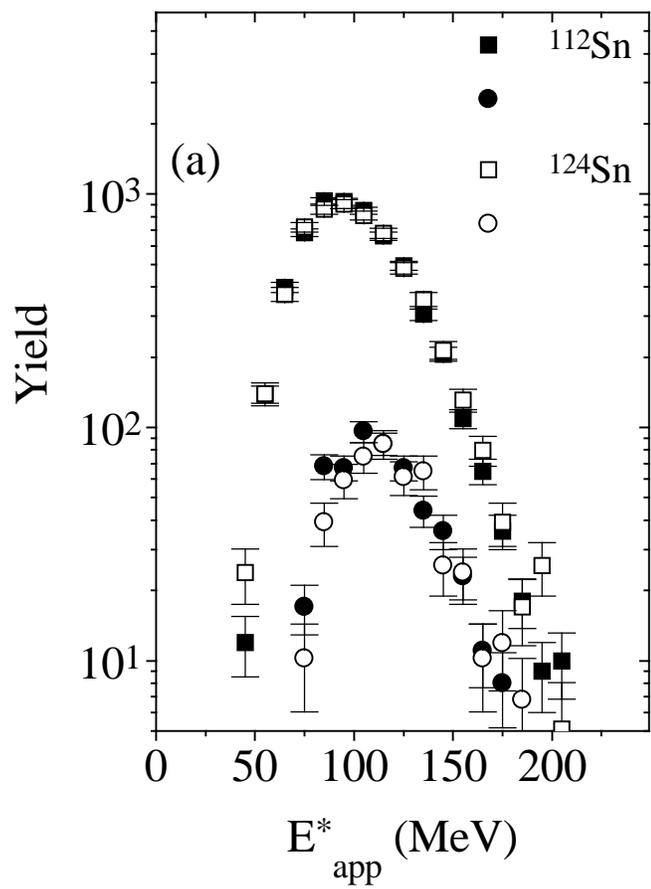
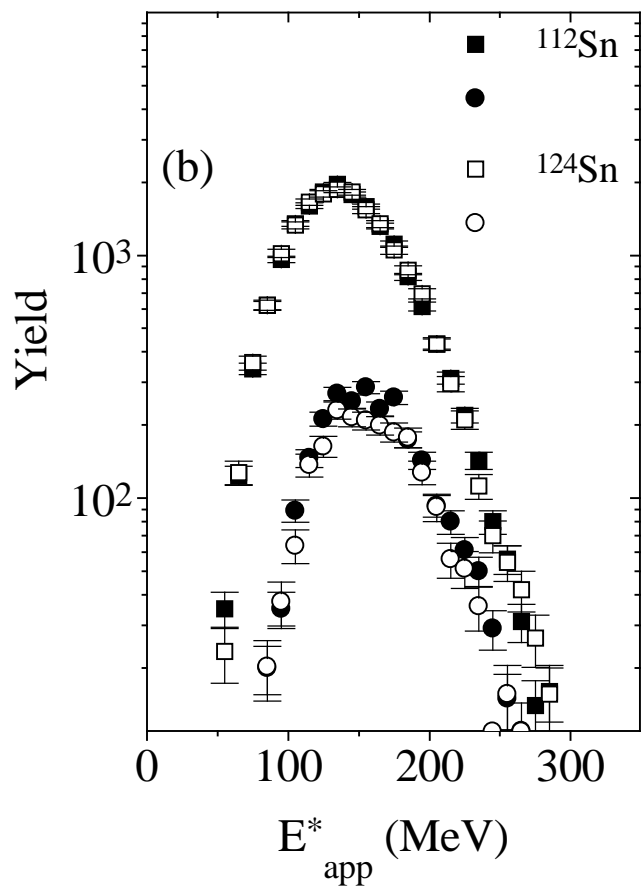
\begin{figure}[!]
\caption{ Experimental distributions of 
the reconstructed apparent excitation energy  of the quasiprojectiles (  
circles - $ Z_{tot}=14 $, squares - $ Z_{tot}= 12-15 $,
solid and open symbols - $^{112}$Sn and $^{124}$Sn target, 
(a) and (b) - projectile energy 
30 and 50 MeV/nucleon ).   }
\label{exex}
\end{figure}

The apparent charged particle excitation energy 
of the quasiprojectile can be reconstructed 
for each projectile fragmentation event from the energy balance 
in the center of mass frame of the quasiprojectile. Thus 

\begin{equation}
 E_{app}^{*}=\sum _{f}(T_{f}^{QP}+\Delta m_{f})-\Delta m_{QP} \hbox{ , }
\label{exqp}
\end{equation}

where $ T_{f}^{QP} $ is the kinetic energy of the fragment in the 
reference frame of the quasiprojectile and $ \Delta m_{f} $ 
and $ \Delta m_{QP} $ are the mass
excesses of the fragment and quasiprojectile, respectively. Emitted 
neutrons are not included in this observable, but for these light 
fragmenting systems neutron emission is not expected 
to dominate. Therefore, $ E_{app}^{*} $ can provide a relative comparison 
of the excitation energy of the fragmenting source at the end 
of the dynamic evolution of the projectile-target system. The distributions of 
the apparent quasiprojectile excitation energies 
reconstructed from fully isotopically resolved events are shown 
in Fig. \ref{exex}. The reconstructed distributions 
for the multifragmentation events with $ Z_{tot}=14 $ are represented 
as circles 
( $^{112}$Sn - solid circles, $^{124}$Sn - open circles ). 
The squares represent the broader set of events with $ Z_{tot}=12-15 $ 
( $^{112}$Sn - solid squares, $^{124}$Sn - open squares ). 
Mean values of the apparent quasiprojectile excitation energies do not 
significantly differ for different targets at the same projectile energy 
and increase with increasing projectile energy ( see Table \ref{table1} ). 
Similar mean values of the excitation energy of reconstructed quasiprojectiles 
for different targets at the same projectile energy suggest similar 
time evolution of the dissipating system. 

The shapes of the excitation energy distributions exhibit 
a strong threshold behavior at low energy and a fast 
decrease in the high energy part which makes them quite narrow. 
They are slightly asymmetric with an excess of yield at high energy. 
The threshold-like behavior may be explained 
by the existence of energy thresholds for the deexcitation channels 
with emission of fragments with $ Z_{f}\leq 5 $. 
Indeed, the low energy threshold behavior does not dramatically change 
with increasing projectile energy. On the other hand, the 
high energy part may be influenced by various factors. The production 
cross section decreases with increasing excitation energy of 
the quasiprojectile. However, it may also be influenced 
by the decrease of the detection efficiency of FAUST  
for multifragmentation events with high multiplicity and large transverse 
momentum. When comparing the shapes of excitation energy distributions 
at different projectile energies, the low energy part is comparable 
at both projectile energies and the range of the excitation energies to which 
the high 
energy part extends increases with projectile energy, thereby increasing the 
widths of quasiprojectile excitation energy distributions.

In order to estimate the influence of target multifragmentation 
on the multiplicity of charged particles, detected at forward angles, 
we used several telescopes positioned at central and backward angles 
for detection of charged particles emitted in coincidence 
with the quasiprojectiles with total charge $ Z_{tot}=12-15 $ where 
all emitted fragments are isotopically identified. Only light charged particles 
( $Z \le 2$ ) were detected at angles between 42.5$^{\circ}$ and 147$^{\circ}$. 
The measured yields of light charged particles were low ( typically 
several tens or a few hundreds of particles per detector ) 
what means the rate not exceeding 0.1 particle detected per isotopically 
resolved quasiprojectile with total charge $ Z_{tot}=12-15 $. 
The yields of charged particles are approximately two times higher for the 
target nucleus $^{112}$Sn than for $^{124}$Sn, which may be explained by 
the lower $N/Z$ ratio of $^{112}$Sn. 
The low multiplicity of coincident charged particles implies that 
the deexcitation of quasitarget is dominated by 
the emission of neutrons, which are not detected in our experiment.  
Although the measured spectra could not be used to estimate 
the slope temperature, we made a rough estimate of the temperature 
of quasitarget from the mean kinetic energy. 
For an ideal Maxwellian spectra, the mean kinetic energy 
of emitted particles above corresponding 
Coulomb barrier is twice the temperature. 
Assuming ideal Maxwellian shape of the measured spectra of protons, 
the values of temperature ranged from 3 to 3.5 MeV 
for both targets and projectile energies.

\subsection*{Simulations}

Experimental distributions of the quasiprojectile observables 
presented above suggest an interplay of nucleon exchange in the early stage, 
leading to partial isospin equilibrium, followed by emission of fragments 
from the highly excited quasiprojectile. 
In order to make more detailed conclusions about the evolution 
of the system, a comparison of experimental observables to the results 
of simulations will be carried out. 
The simulation will include a model description of the reaction dynamics 
and a software replica of the FAUST multidetector array ( filter routine ).  
For the model description to be considered as adequate we require   
both $ \Delta A $ and $ E_{app}^{*} $ to be optimally reproduced 
for different targets, projectile energies and subsets of data. 

The basic assumption on which the simulation is based is the possibility 
to decompose the collision into two stages. 
In the early stage of the collision hot quasiprojectiles are created 
which then deexcite by the statistical decay. To describe the 
production of excited quasiprojectiles we used the Monte Carlo code 
of Tassan-Got et al. \cite{tassangot}. This code implements a 
version of the model of deep inelastic transfer suitable for Monte Carlo 
simulations. 
For each event, the system evolution is determined by random transfers 
of nucleons between the projectile and target through 
an open window between the nuclei. For each transfer, the internal 
and relative velocities are coupled. Even though the traditional domain 
of deep inelastic transfer lies at energies below 
20 MeV/nucleon, the comparison of the calculated and experimental fragment 
energy and mass distributions seems 
to give reasonable agreement up to projectile energies 50 MeV/nucleon 
when the effect of fragment deexcitation is included \cite{tassangot}. 
The parameters of the model used in this work are identical 
to the parameters used in the original work \cite{tassangot}. 
The number of events generated at a given angular momentum was proportional 
to the geometrical cross section for a given partial wave. 
Mean values and widths of the quasiprojectile
excitation energy, mass and charge distributions of generated events with
intrinsic excitation of the quasiprojectile higher than 35 MeV 
and $Z_{QP}$ = 12 - 15 are given in Table \ref{table2}. 

\begin{table}[tbp]
\caption{ Mean values and widths of quasiprojectile
excitation energy, mass and charge distributions 
simulated using the model of deep 
inelastic transfer \cite{tassangot}. Only excited quasiprojectiles  
with $Z_{QP}$ = 12 - 15 and intrinsic excitation 
higher than 35 MeV are included. }
\label{table2}

{\centering \begin{tabular}{cccccccc}
\hline 
\hline 
$ E_{proj} $&
Target&
 $ <A_{QP}> $ &
 $ \sigma _{A_{QP}} $ &
$ <Z_{QP}> $ &
$ \sigma _{Z_{QP}} $ &
$ <E^{*}_{QP}> $&
$ \sigma _{E^{*}_{QP}} $\\
(MeV/nucleon)&
&
&
&
&
&
(MeV)&
(MeV)\\
\hline 
30&
$ ^{112} $Sn&
28.01&
2.36&
13.54&
1.20&
95.5&
42.7\\
&
$ ^{124} $Sn&
28.46&
2.15&
13.25&
1.03&
95.1&
44.2\\
50&
$ ^{112} $Sn&
27.91&
2.23&
13.64&
1.19&
146.2&
78.5\\
&
$ ^{124} $Sn&
28.19&
2.09&
13.34&
1.04&
143.2&
79.5\\
\hline 
\hline 
\end{tabular}\par}

\end{table}

At both projectile energies, the number of neutrons transferred 
from the target to the projectile increases with the neutron number of the 
target. A heavier target with a larger neutron 
number also causes stronger proton flow from the projectile to the target. 
This behavior is caused by an evolution towards isospin equilibrium between 
projectile and target and is in qualitative agreement with experimental trends.  
Mean excitation energies of the simulated quasiprojectiles  
are comparable to the experimental
values in Table 
\ref{table1} and exhibit the same trends for given beam energies and 
target nuclei. The widths of simulated inclusive excitation energy 
distributions are larger than the experimental data. 

The simulated mean excitation energies of the target are slightly higher 
than 
1 MeV/nucleon for the projectile energy of 30 MeV/nucleon. At 50 MeV/nucleon 
they reach  1.5 MeV/nucleon. The corresponding temperatures 
obtained using the well known formula $ T = (E^{*}/\tilde{a})^{1/2} $ 
are in reasonable agreement with the estimated temperature of the target,  
when using the asymptotic value of the level density parameter 
$ \tilde{a} = A/9 $. At these temperatures, between 3 and 3.5 MeV, 
the emission of neutrons 
may be expected to be a dominating deexcitation channel for 
nuclei with mass and charge close to the target. 

In general, the concept of deep inelastic transfer reasonably describes 
the early stage of the collisions investigated 
in the present experiment. When combined with a realistic deexcitation 
model, it may provide a good general description of the reaction 
mechanism. Since the mean value of excitation energy 
per nucleon is well above 3 MeV/nucleon for both projectile energies, 
we simulated the deexcitation of the highly excited quasiprojectile using 
the statistical model of multifragmentation ( SMM ) \cite{bondorf}.
Macrocanonical partitions of the hot fragments were generated 
for individual events. 
For the hot fragments emitted from the quasiprojectile, a multiparticle Coulomb 
tracking was applied. The final partition of cold fragments was obtained by 
deexcitation of the hot fragments via Fermi decay and particle emission. 
The quasiprojectile event sequences generated by the DIT code 
of Tassan-Got \cite{tassangot} have been used as the input of SMM simulations. 
The deexcitation of the excited quasitarget was not taken into account as 
a contributing source of the charged particles at forward angles. 

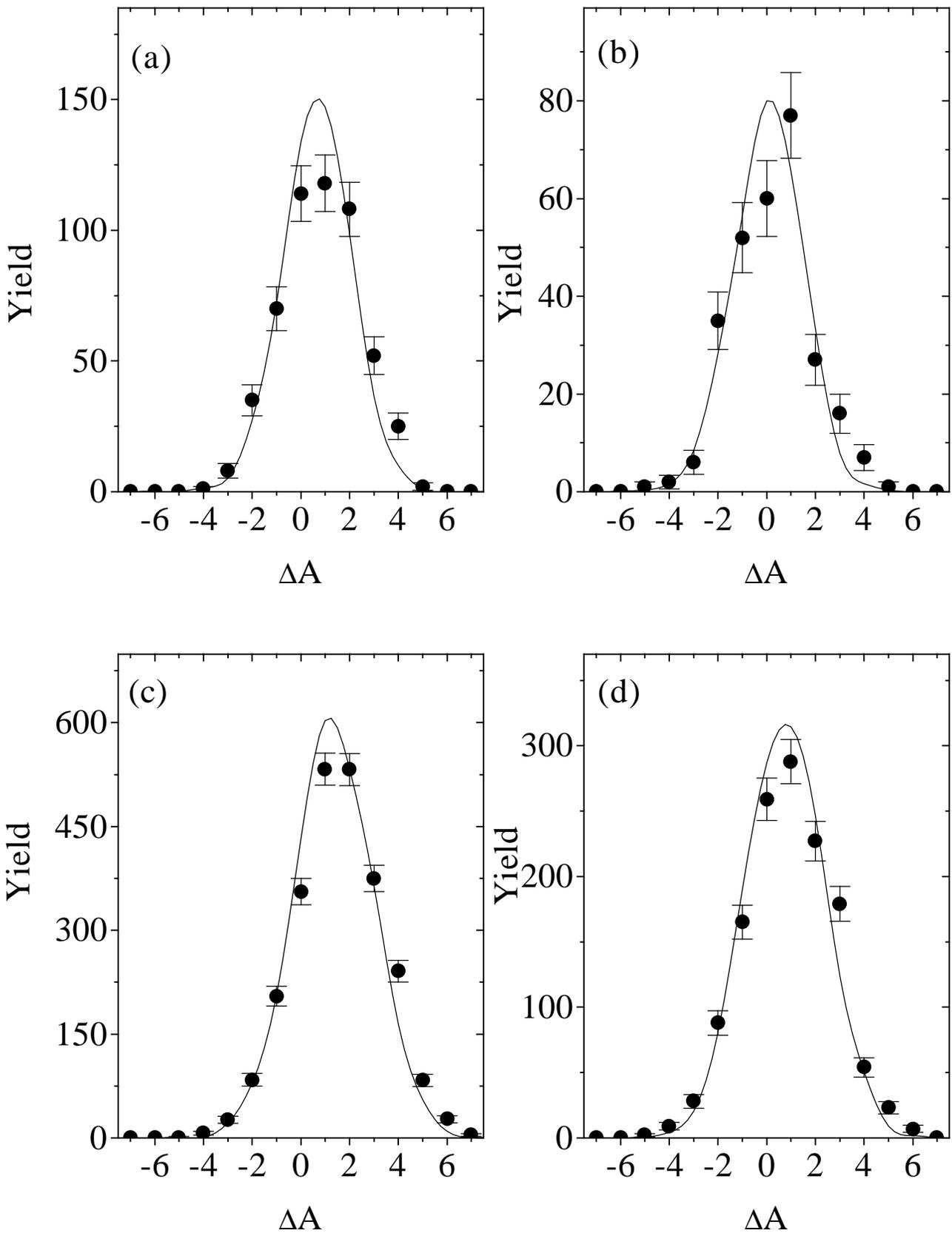
\begin{figure}[!]
\caption{ Experimental ( solid circles ) and 
simulated ( solid lines ) mass change distributions 
for the fully isotopically resolved quasiprojectiles with $ Z_{tot}=14 $, 
(a) - $^{28}$Si(30MeV/nucleon) + $^{112}$Sn, 
(b) - $^{28}$Si(30MeV/nucleon) + $^{124}$Sn, 
(c) - $^{28}$Si(50MeV/nucleon) + $^{112}$Sn, 
(d) - $^{28}$Si(50MeV/nucleon) + $^{124}$Sn.
For details of simulation see text. }
\label{munpsim}
\end{figure}

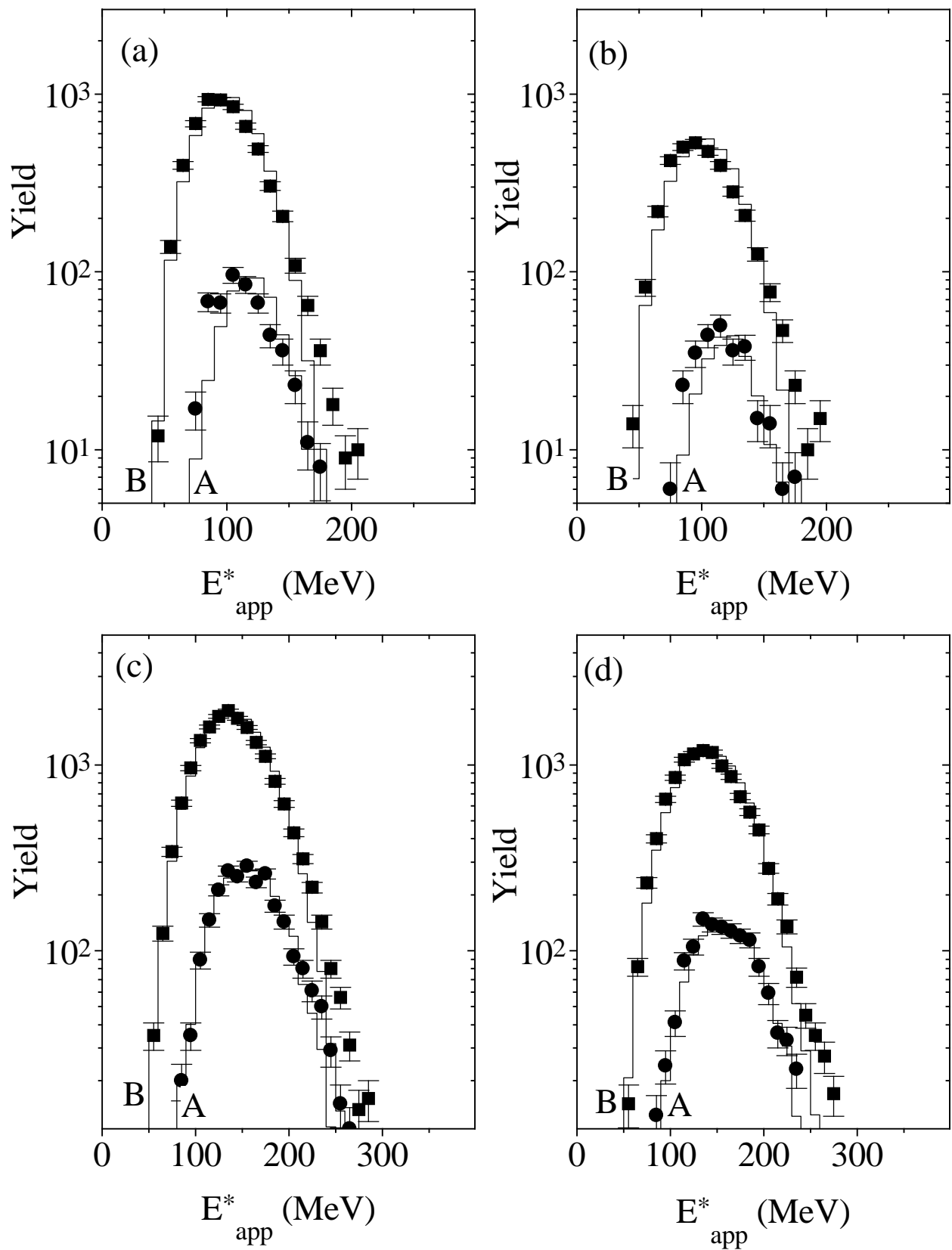
\begin{figure}[!]
\caption{ Distributions of the reconstructed apparent excitation energies 
of the quasiprojectiles. Symbols mean experimental distributions of  
the set of fully isotopically resolved quasiprojectiles  
with $ Z_{tot}=14 $ ( solid circles )  
and $ Z_{tot}=12-15 $ ( solid squares ). 
Solid histograms labeled as A and B mean simulated distributions
for $ Z_{tot}=14 $ and $ Z_{tot}=12-15 $, 
(a) - $^{28}$Si(30MeV/nucleon) + $^{112}$Sn, 
(b) - $^{28}$Si(30MeV/nucleon) + $^{124}$Sn, 
(c) - $^{28}$Si(50MeV/nucleon) + $^{112}$Sn, 
(d) - $^{28}$Si(50MeV/nucleon) + $^{124}$Sn.
For details of simulations see text. }
\label{exsim}
\end{figure}

To mimic the experimental selection criteria, we employed restrictions 
on the kinetic and excitation energy of the simulated quasiprojectiles. 
Only those events where the quasiprojectiles satisfied the relation 
$ \sin {\theta _{lim}}=\frac{<p_{f}^{QP}>}{<p_{f||}^{Lab}>} 
\approx \sqrt{\frac{E^{*}_{QP}}{E_{kinQP}^{Lab}}}\leq 0.6 $ 
and had intrinsic excitation energy greater than 35 MeV 
were used as an input to the SMM calculations. 
This relation rejects the events with fragments emitted 
outside of the acceptance of our detector setup. 
The variable $ <p_{f}^{QP}> $ is 
the mean fragment momentum in the quasiprojectile center of mass 
frame, $ <p_{f||}^{Lab}> $ is the mean value of the component of 
fragment momentum in the laboratory frame parallel to the beam axis 
and $ E_{kinQP}^{Lab} $ is 
the kinetic energy of quasiprojectile in the laboratory frame.  
The initial nuclear density of the fragmenting quasiprojectile was equal 
to the equilibrium nuclear density. The SMM events with 
all fragments having $ Z_{f}\leq 5 $ 
were filtered by the FAUST software replica, which simulates the geometrical 
coverage of FAUST and the energy thresholds of the telescopes 
for a given fragment mass and charge. The results of the simulation are
shown in \hbox{Figs. \ref{munpsim},\ref{exsim}}. 
The simulated distributions of the mass change 
for fully isotopically resolved events with $ Z_{tot}=14 $ ( solid lines ) 
are plotted in Fig. \ref{munpsim} ( solid circles represent experimental data )
normalized to the number of experimental events 
with $ Z_{tot}=14 $.
The agreement of the experimental and simulated distributions of 
the mass change 
is quite good. In Fig. \ref{exsim} the simulated distributions
of the apparent quasiprojectile excitation energy are shown for 
both $ Z_{tot}=14 $ and $ Z_{tot}=12-15 $ 
( solid histograms labeled as A and B respectively ) 
along with the experimental data ( solid circles and squares respectively ). 
The simulated data have been 
normalized to the sum of experimental events with $ Z_{tot}=12-15 $. The 
agreement of the simulated and experimental apparent quasiprojectile 
excitation energy
distributions with both $ Z_{tot}=12-15 $ and $ Z_{tot}=14 $ is 
quite good. 
The onset of multifragmentation into channels with 
$ Z_{f} \leq 5 $ in the low energy part is described with good precision 
for both sets of data $ Z_{tot}=12-15 $ and $ Z_{tot}=14 $. 

The mean values of the mass and charge of the quasiprojectiles with 
$ Z_{tot}=12-15 $ obtained from the simulation are comparable with 
the values, obtained from experimental mass and charge distributions 
( see Table \ref{table_az} ). The simulation is able to reproduce 
the trends of the nucleon exchange also for the broader set of contributing 
events which are not taken into account in the analysis of $ \Delta A $. 

\begin{table}[tbp]
\caption{ 
Mean values of the mass and charge of the reconstructed quasiprojectiles 
with $Z_{tot}$ = 12 - 15. Experimental values are compared 
to the results of simulation. For details of simulation see text. 
}
\label{table_az}

{\centering \begin{tabular}{cccccc}
\hline 
\hline 
$ E_{proj} $ &
Target&
\multicolumn{2}{c}{ $ <Z_{QP}> $ }&
\multicolumn{2}{c}{ $ <A_{QP}> $ }\\
(MeV/nucleon)&
&
Exp.&
Sim.&
Exp.&
Sim.\\
\hline 
30 &
$ ^{112} $Sn&
12.55$ \pm  $0.26&
12.51$ \pm  $0.11&
24.36 $ \pm  $0.56&
24.65$ \pm  $0.25\\
&
$ ^{124} $Sn&
12.50$ \pm  $0.33&
12.39$ \pm  $0.11&
25.00 $ \pm  $0.71&
25.15$ \pm  $0.25\\
50 &
$ ^{112} $Sn&
12.74$ \pm  $0.17&
12.72$ \pm  $0.10 &
24.33 $ \pm  $0.34&
24.68$ \pm  $0.24\\
&
$ ^{124} $Sn&
12.67$ \pm  $0.19&
12.61$ \pm  $0.10&
24.83 $ \pm  $0.39&
24.96$ \pm  $0.24\\
\hline 
\hline 
\end{tabular}\par}

\end{table}

The overall agreement in Figs. \ref{munpsim},\ref{exsim} 
and Table \ref{table_az} 
shows that a combination 
of the concepts of deep inelastic transfer and statistical multifragmentation 
satisfactorily describes the data. 
The influence of the target neutron number and beam energy is reproduced 
correctly not only on average, but even for different subsets of data. 
Using a backtracing procedure we estimated the range of contributing 
angular momenta in the simulated data. 
The mean values are 186 and 203 $ \hbar  $ for $^{112}$Sn 
and $^{124}$Sn 
targets at projectile energy 30 MeV/nucleon and 243 and 263 $ \hbar  $ 
for 50 MeV/nucleon, respectively.  When converting angular momentum 
to impact parameter, the mean values are 5.5 and 5.6 fm 
for $^{112}$Sn and $^{124}$Sn targets at projectile energy 30 MeV/nucleon 
and 5.9 and 6.0 fm at the higher projectile energy. 
Estimated mean values are well below the contact ( grazing ) configuration 
for both reactions, which can be roughly estimated with  
$ R_{12}=r_{0}(A_{P}^{\frac{1}{3}}+A_{T}^{\frac{1}{3}})+1\hbox{ fm} $. 
When taking $ r_{0}=1.12 \hbox{ fm} $ , corresponding to half-density radii, 
the $ R_{12} $ equals 9.8 and 10.0 fm for $^{112}$Sn and 
$^{124}$Sn, respectively. 
The estimated range of contributing impact parameters corresponds 
to peripheral collisions. 

Using the backtracing procedure, we estimated the mean multiplicity of 
neutrons emitted from the quasiprojectile. Mean values of the multiplicity of 
emitted neutrons are 0.9 and 1.2 for $^{112}$Sn 
and $^{124}$Sn targets at projectile energy 30 MeV/nucleon 
and 1.4 and 1.7 for $^{112}$Sn 
and $^{124}$Sn targets at projectile energy 50 MeV/nucleon, respectively. 
The mean values of the multiplicity of 
emitted neutrons increase with increasing projectile energy 
while the effect of target neutron excess is relatively weak.  
Using the estimated multiplicities of emitted neutrons, we determined 
the mean values of the $N/Z$ ratio of the excited quasiprojectiles 
to be 1.04 and 1.12 for $^{112}$Sn 
and $^{124}$Sn targets at projectile energy 30 MeV/nucleon 
and 1.03 and 1.11 for $^{112}$Sn 
and $^{124}$Sn targets at projectile energy 50 MeV/nucleon, respectively. 
The mean values of the $N/Z$ ratio for the different projectile energies 
only slightly differ. The effect of target isospin is significant 
and is similar 
at both projectile energies. The estimated values of the $N/Z$ ratio 
of the hot quasiprojectile are consistent with the simulated mean values 
of the mass and charge of excited quasiprojectiles given in Table \ref{table2}. 

Even if the projectile energies are relatively high, the
data do not appear to be strongly influenced by preequilibrium 
emission. For the protons, which are primary candidates for 
preequilibrium emission among particles observed in the present 
experiment, we determined their momenta in the quasiprojectile 
frame and constructed two dimensional plots of the momentum component 
parallel to the quasiprojectile direction versus the momentum 
of quasiprojectile in the lab frame. In the experimental distributions for the 
 $^{112}$Sn target, two different sources could be 
identified, a stronger one in the forward hemisphere and a weaker one in 
the backward hemisphere of the quasiprojectile. The experimental distributions 
for the $^{124}$Sn target consisted only of the particles in forward 
hemisphere, which was fully compatible with the forward source in the 
previous case. This feature is unique for protons and does not exist 
in the case of heavier fragments. 
According to the conclusions of work \cite{charity}, the protons in forward 
hemisphere of the quasiprojectile are emitted from the quasiprojectile during 
multifragmentation and are later shifted forward by the Coulomb field of the 
target because of the high charge to mass ratio compared to other fragments. 
The systematics of Coulomb shifts observed in our data tracks well with the 
results of ref. \cite{charity}. Thus, in the case of $^{112}$Sn target, the 
protons in the backward hemisphere can be attributed to preequilibrium emission. 
The absence of such a source in the case of $^{124}$Sn target can be explained 
by the emission of preequilibrium neutrons from this more neutron rich 
system which are not detected in our experiment. From the event rate 
in this component, we estimated the multiplicity of preequilibrium 
protons accompanying multifragmentation of the fully isotopically 
resolved quasiprojectiles with $Z_{tot}=12-15$ as 0.2$\pm$0.1 for the 
projectile energy 30 MeV/nucleon and 0.3$\pm$0.1 for the projectile energy 
50 MeV/nucleon. Such rates are quite moderate and the physical 
picture used by the simulation remains valid. 

To summarize this section, we presented a unique set of isotopically resolved
projectile multifragmentation data and determined the dominant
mechanism of nucleon exchange, but not without using model assumptions 
about deexcitation of the excited quasiprojectile. In the next section 
we will discuss the deexcitation of the quasiprojectile in detail.

\section*{Quasiprojectile multifragmentation}

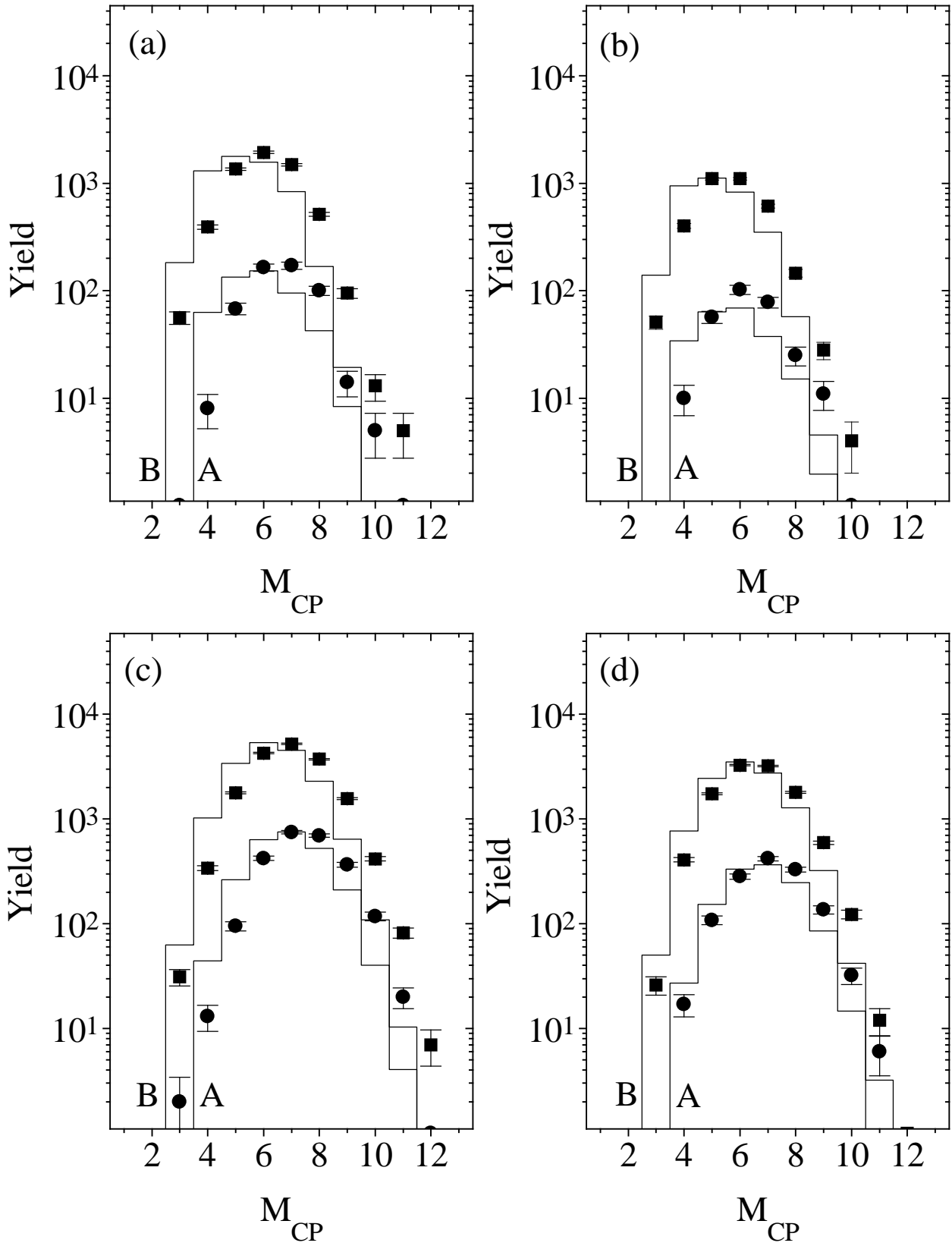
\begin{figure}[!]
\caption{ Experimental multiplicity distributions of charged particles 
emitted from the fully isotopically resolved quasiprojectiles with 
$ Z_{tot}=14 $ ( solid circles ) and $ Z_{tot}=12-15 $ ( solid squares ) 
and the results of simulation for the same sets 
of data ( histograms labeled as A and B ), 
(a) - $^{28}$Si(30MeV/nucleon) + $^{112}$Sn, 
(b) - $^{28}$Si(30MeV/nucleon) + $^{124}$Sn, 
(c) - $^{28}$Si(50MeV/nucleon) + $^{112}$Sn, 
(d) - $^{28}$Si(50MeV/nucleon) + $^{124}$Sn.
For details of simulations see text. }
\label{mcp}
\end{figure}

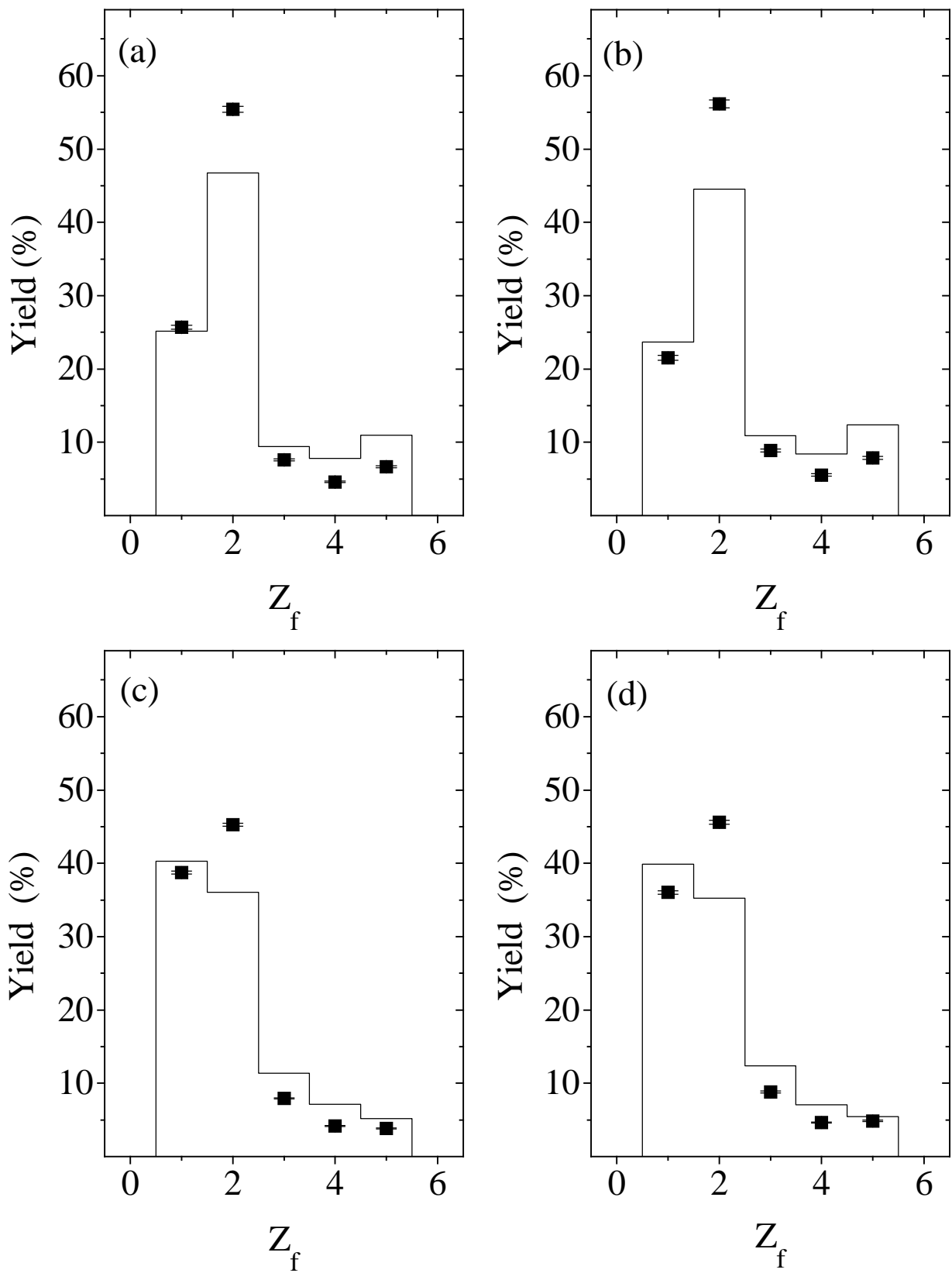
\begin{figure}[!]
\caption{ Experimental ( solid squares ) 
and simulated ( solid histograms ) charge distributions of fragments
emitted from the fully isotopically resolved quasiprojectiles with 
$ Z_{tot}=12-15 $, 
(a) - $^{28}$Si(30MeV/nucleon) + $^{112}$Sn, 
(b) - $^{28}$Si(30MeV/nucleon) + $^{124}$Sn, 
(c) - $^{28}$Si(50MeV/nucleon) + $^{112}$Sn, 
(d) - $^{28}$Si(50MeV/nucleon) + $^{124}$Sn.
For details of simulations see text.  }
\label{zdist}
\end{figure}

In order to obtain a fully isotopically resolved event with $ Z_{f}\leq 5 $,
the quasiprojectiles with $ Z_{tot}=12-15 $ have to disintegrate 
into at least three charged particles. As already shown in
Fig. \ref{exsim}, the simulation is capable of correctly describing the onset 
of this fragmentation mode and the overall quasiprojectile excitation energy 
distribution for the quasiprojectiles with the charge close to the 
charge of the projectile. The experimental distributions of charged particle 
multiplicity are presented in \hbox{Fig. \ref{mcp}} for isotopically 
resolved data with $ Z_{tot}=14 $ ( solid circles ) and  $ Z_{tot}=12-15 $ 
( solid squares ). The simulations are presented as histograms labeled 
as A ( $ Z_{tot}=14 $ ) and B ( $ Z_{tot}=12-15 $ ). The calculations 
are normalized to the experimental data by the sum of isotopically resolved 
quasiprojectiles with $ Z_{tot}=12-15 $. The simulated 
data in Fig. \ref{mcp} show reasonable overall agreement with the results 
of experiment. The simulated distributions are shifted to somewhat 
lower values of multiplicity. Table \ref{table3} presents the mean 
values of the fragment multiplicity and fragment  
charge for both experiment and simulation. The simulated mean fragment 
multiplicities are smaller than the experimental ones. The difference ranges 
from 0.2 to 0.6 and is slightly larger in the case of $^{112}$Sn. The simulated 
mean values ( see Table \ref{table3} ) of the fragment charge are larger 
than the experimental ones, thus counterbalancing a smaller fragment number. 
The fragment charge yields ( see Fig. \ref{zdist} ) show analogous yields 
of fragments with $ Z_{f}=1 $, the simulated yields  
of fragments with $ Z_{f}=2 $ are smaller than the experimental ones 
by about 10 \% and the simulated yields of heavier fragments are
higher than the experimental ones. 
Higher experimental yield of $ \alpha  $-particles may be influenced by the 
existence of pre-formed $ \alpha  $-clusters in the projectile 
nucleus $^{28}$Si. The data presented applies to isotopically 
resolved events with $ Z_{tot}=12-15 $. Similar distributions 
for subsets of data with $ Z_{tot}=14 $ give identical results.

\begin{table}[tbp]
\caption{ 
Mean values of the multiplicity, charge, 
and $N/Z$ ratio of the charged fragments, 
emitted in events where the quasiprojectiles with $Z_{tot}$ = 12 - 15  
were fully isotopically resolved. Experimental values are compared 
to the results of simulation. For details of simulation see text. 
}
\label{table3}

{\centering \begin{tabular}{cccccccc}
\hline 
\hline 
$ E_{proj} $ &
Target&
\multicolumn{2}{c}{ $ <M_{CP}> $ }&
\multicolumn{2}{c}{ $ <Z_{f}> $ }&
\multicolumn{2}{c}{ $ <N/Z> $ }\\
(MeV/nucleon)&
&
Exp.&
Sim.&
Exp.&
Sim.&
Exp.&
Sim.\\
\hline 
30 &
$ ^{112} $Sn&
5.95$ \pm  $0.11&
5.37$ \pm  $0.04&
2.11$ \pm  $0.02&
2.33$ \pm  $0.01&
0.95$ \pm  $0.02&
0.97$ \pm  $0.01\\
&
$ ^{124} $Sn&
5.63$ \pm  $0.14&
5.14$ \pm  $0.04&
2.22$ \pm  $0.02&
2.41$ \pm  $0.01&
1.01$ \pm  $0.02&
1.03$ \pm  $0.01\\
50 &
$ ^{112} $Sn&
6.74$ \pm  $0.08&
6.33$ \pm  $0.04&
1.89$ \pm  $0.01&
2.01$ \pm  $0.01 &
0.91$ \pm  $0.01&
0.94$ \pm  $0.01\\
&
$ ^{124} $Sn&
6.43$ \pm  $0.09&
6.21$ \pm  $0.04&
1.97$ \pm  $0.01&
2.03$ \pm  $0.01&
0.96$ \pm  $0.01&
0.98$ \pm  $0.01\\
\hline 
\hline 
\end{tabular} \par}

\end{table}

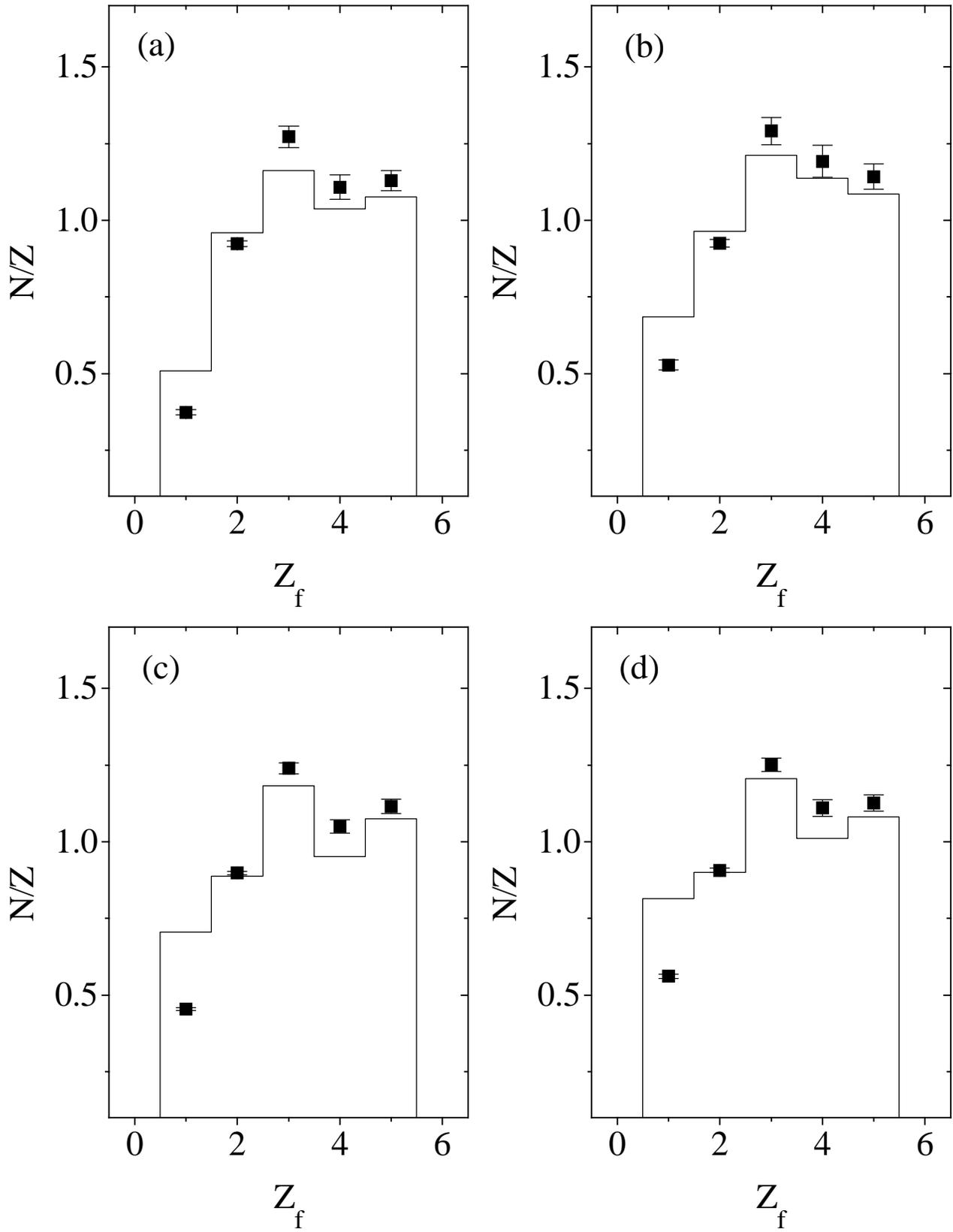
\begin{figure}[!]
\caption{ Experimental ( solid squares ) and simulated ( solid histogram ) 
dependences of mean fragment $N/Z$ ratio on the charge of fragments  
emitted from the fully isotopically resolved quasiprojectiles 
with $ Z_{tot}=12-15 $,  
(a) - $^{28}$Si(30MeV/nucleon) + $^{112}$Sn, 
(b) - $^{28}$Si(30MeV/nucleon) + $^{124}$Sn, 
(c) - $^{28}$Si(50MeV/nucleon) + $^{112}$Sn, 
(d) - $^{28}$Si(50MeV/nucleon) + $^{124}$Sn.
For details of simulations see text. 
}
\label{nzfrag}
\end{figure}

Additional understanding of quasiprojectile deexcitation 
may be obtained from the study of isotopic degrees of freedom. 
The overall values of the $N/Z$ of the quasiprojectile are 
similar for the experiment 
and the simulations ( see Table \ref{table3} ). The results of simulation are 
slightly higher in all cases but the difference is within 
the statistical errors. The situation 
is significantly different when investigating the fragments 
of different charges independently. 
In Fig. \ref{nzfrag}, we present average $N/Z$ ratios for fragments
with different charges. The data presented applies to isotopically 
resolved events with $ Z_{tot}=12-15 $. The results 
for subsets of data with $ Z_{tot}=14 $ are 
practically identical. Experimental $N/Z$ ratios show 
an excess of neutron rich fragments with $ Z_{f}\geq 3 $ 
relative to the simulation, counterbalanced by stronger
dominance of protons among the fragments with $ Z_{f}=1 $. This may point
out a higher decay probability of the excited neutron deficient 
quasiprojectiles 
or hot fragments for the channels with emission of stable charged particles 
like protons and $ \alpha  $-particles.
Alternatively, the relative excess of protons may be 
caused by preequilibrium emission, especially in the case of 
less neutron rich target $^{112}$Sn. 
When comparing the sensitivity of experimental $N/Z$ ratios 
to the neutron content of the target at given projectile energy, 
the $N/Z$ ratios of fragments with $ Z_{f}=1 $
and $ Z_{f}=4 $ show the highest sensitivity. This trend was reported in
our previous study where a broader set of data was presented \cite{laforest}. 

For the case of $^{8}$Li, which could be influenced by an admixture 
of the two $\alpha$-particle decay of short-lived $^{8}$Be, we compared the 
experimental and simulated values of the isotopic ratio 
Y($^{8}$Li)/Y($^{7}$Li) for different bins of the isospin of the 
quasiprojectile. Detection of $^{8}$Be was a priori excluded in the 
simulation. We found no significant deviations between experimental 
data and simulation, which allows 
us to conclude that the admixture of $^{8}$Be in the yield of $^{8}$Li 
does not dramatically influence the results of our analysis. 

In summary, the overall description of the experimental data on charged 
particle
multiplicity, charge distributions and isotopic ratios may be considered as
reasonable in general. The remaining minor inconsistencies may be attributed 
to the limitations in the model description of quasiprojectile deexcitation 
and/or to the influence of preequilibrium emission. 
These inconsistencies, however, do not influence conclusions concerning
the mechanism of nucleon exchange given in previous section.

\section*{Summary}

Using the FAUST detector array we obtained a set of fully isotopically 
resolved projectile multifragmentation events ( $ Z_{f}\leq 5 $ ) 
from the reactions 
$^{28}$Si+$^{112,124}$Sn at projectile energies 30 and 50 MeV/nucleon.  
We have been able to reconstruct the mass, charge
and dynamic observables of the excited quasiprojectile and to 
study the nucleon exchange between projectile and target. 
The reconstructed velocity distributions of the emitting source 
have been fitted using one Gaussian source. 
Thus admixtures from a midvelocity source may be excluded. 
At a given projectile energy, we observed an influence of the target 
isospin on the mass change 
of the reconstructed quasiprojectiles that have the charge 
of the beam \hbox{( $ Z_{tot}=14 $ )}. 
However, we observed no significant influence of the target isospin on the 
apparent excitation energy distribution. Reactions with a heavier target 
isotope result in lower average mass change. 
This may be seen as evidence for partial equilibration 
of isospin in the early stage of the reaction. In  
the reactions with the same target the mass change 
increases with increasing projectile energy. 
This corresponds to a shift in the distributions of apparent  
quasiprojectile excitation energies toward higher values. 
The influence of the target isospin and of the projectile energy 
on the neutron content of the reconstructed quasiprojectiles can be explained 
by a two stage model consisting of nucleon exchange in the early stage of 
collision followed by deexcitation of the quasiprojectile. 

The experimental observables 
for different subsets of data were reproduced using a simulation
using the concept of deep inelastic transfer in the early stage followed by 
quasiprojectile multifragmentation and sequential decay of the hot fragments. 
The deexcitation of the excited 
quasitarget and the preequilibrium emission were not taken into  
account in our simulation. 
Distributions of the mass change 
and apparent excitation energy of reconstructed quasiprojectiles 
have been reproduced with good overall agreement. 
The charged fragment multiplicities, charge distributions and $N/Z$ ratios
for different fragment charges imply lower experimental survival probability 
of neutron deficient fragments towards decay into stable light charged 
particles than predicted by the simulation. 
We observed a maximum of the sensitivity of the $N/Z$ ratios to the target 
isospin for the fragment charges $ Z_{f}=1 $ and $ Z_{f}=4 $. 
The contributing range of impact parameters was estimated 
by backtracing the simulated data ( $ b= 5 - 7 \hbox{ fm} $ ) 
indicating that the collisions may be considered as nearly peripheral. 
Observables related to target multifragmentation and 
preequilibrium emission imply that neither of the processes causes 
significant distortion of the physical picture used in the simulation. 
The backtracing of simulated data allowed an estimation of the 
multiplicity of neutrons emitted from the quasiprojectile 
in the deexcitation stage. The estimated neutron multiplicities allowed 
further determination of the corresponding level of isospin equilibration 
between projectile and target during the nucleon exchange stage, which 
strongly depends on target isospin. 

The present work shows that deep inelastic transfer is the dominant 
production mechanism of highly excited quasiprojectiles in peripheral 
collisions in the Fermi energy domain and that such collisions are suitable 
for detailed studies of thermal multifragmentation.

\begin{acknowledgments}

The authors wish to thank the Cyclotron Institute staff for the excellent
beam quality. This work was supported 
in part by the NSF through grant No. PHY-9457376, 
the Robert A. Welch Foundation through grant No. A-1266, and 
the Department of Energy through grant No. DE-FG03-93ER40773.
M. V. was partially supported through grant VEGA-2/5121/98. 

\end{acknowledgments}

\newpage 

\section*{ }

\small

\end{document}